\documentclass[12pt]{article}
\usepackage{epsfig}
\newcommand{\be}{\begin{equation}}
\newcommand{\ee}{\end{equation}}
\newcommand{\ba}{\begin{eqnarray}}
\newcommand{\ea}{\end{eqnarray}}
\newcommand{\tr}{{\rm Tr\,}}
\newcommand{\ii}{{\rm i}}

\newcommand{\ex}{{\rm e}}
\newcommand{\nn}{\nonumber}

\newcommand{\bfx}{{\bf x}}

\newcommand{\bmu}{\bar{\mu}}

\newcommand{\op}{{\cal O}}
\newcommand{\eq}{Eq.~}
\newcommand{\eqs}{Eqs.~}
\newcommand{\fig}{Fig.~}
\newcommand{\RR}{{\rm I\kern -.2em  R}}
\def\lsi{\raise0.3ex\hbox{$<$\kern-0.75em\raise-1.1ex\hbox{$\sim$}}}
\def\gsi{\raise0.3ex\hbox{$>$\kern-0.75em\raise-1.1ex\hbox{$\sim$}}}
\newcommand{\lsim}{\mathop{\lsi}}

\def\none               {\multicolumn{2}{c|}{--}}

\begin{document}

\begin{titlepage}
\begin{flushright}
MIT-CTP-3270\\
CERN-TH/2002-102
\end{flushright}
\begin{centering}
\vfill

{\bf \Large The QCD phase diagram for small densities\\
from imaginary chemical potential}

\vspace{0.8cm}

Philippe~de Forcrand$^{\rm a,b}$ and
Owe~Philipsen$^{\rm c}$

\vspace{0.3cm}

{\em $^{\rm a}$
Institut f\"ur Theoretische Physik,
ETH Z\"urich,
CH-8093 Z\"urich, 
Switzerland\\}
{\em $^{\rm b}$
Theory Division, CERN, CH-1211 Geneva 23,
Switzerland\\}
{\em $^{\rm c}$
Center for Theoretical Physics, Massachusetts Institute of Technology,\\
Cambridge, MA 02139-4307, USA}

\vspace*{0.7cm}

\begin{abstract}
We present results on the QCD phase diagram for $\mu_B\leq \pi T$.
Our simulations are performed with an imaginary chemical potential $\mu=i\mu_I$
for which the fermion determinant is positive.
On an $8^3\times 4$ lattice with 2 flavors of staggered quarks,
we map out the phase diagram and
identify the (pseudo-)critical temperature $T_c(\mu_I)$.
For $\mu_I/T \leq \pi/3$, this is an analytic function,
whose Taylor expansion is found to converge rapidly,
with truncation errors far smaller than statistical ones.
The truncated series may then be continued to real $\mu$,
yielding the corresponding phase diagram for $\mu_B\lsim 500$ MeV.
This approach provides control over systematics and avoids reweighting.
We compare it with other recent work.
\end{abstract}

\end{centering}
\noindent
\vfill
\end{titlepage}

\section{Introduction}

A prime goal of heavy ion collision experiments at SPS, LHC (CERN) and RHIC (Brookhaven)  
is to probe the transition\footnote{In this paper the term "transition" 
is used to label a change in
dynamics, whether it proceeds by a smooth crossover or by a
first or second order phase transition. For the last two, we use explicitly the
term "phase transition".}
from hadronic matter to a quark gluon plasma at
high temperatures and small baryon density. 
Because QCD is strongly interacting,
the only first principles method to predict the phase diagram is by means 
of lattice simulations.
On the other hand, at finite densities the fermion determinant is complex,
thus prohibiting standard Monte Carlo sampling of the path integral in what is 
known as the `sign problem'. Overviews with references to various studies
of this problem may be found in \cite{rev}.

Recently, a first $(\mu,T)$ phase diagram for 2+1-flavor QCD has been presented 
\cite{fk2} by using a two-dimensional generalization \cite{fk1} of the Glasgow
reweighting method \cite{rew}. For such methods cancellations in the reweighting factor 
occur generically at large volumes and/or chemical potentials, 
and it remains to be seen how well this approach allows
for infinite volume and continuum extrapolations.
Another work \cite{hk} attempts to improve on this aspect by
expanding the product of reweighting factor and observable in powers of $(\mu/T)$, 
and computing the coefficients through second order.
The calculation is then equivalent to computing susceptibilities at $\mu=0$ 
\cite{susc}, which is possible on any volume. The $\mu/T$-range of 
applicability is limited by either the $\mu/T$-value of the
expected critical point marking the onset of a phase transition,
where susceptibilities become non-analytic, or the radius of convergence of the 
expansion, whichever is lower. 
However, a priori nothing is known about the convergence of the
expansion or the error introduced by its truncation.
Moreover, a problem common to all reweighting approaches is 
that so far it has been impossible
to assess the overlap of reweighted ensembles with the original. 
Hence, the results obtained in \cite{fk2,hk} rest on as yet 
uncontrolled approximations.

In this paper we attempt to develop a controllable alternative approach, avoiding 
reweighting in $\mu$ altogether.
This is achieved by simulating with imaginary chemical potential, 
where there is no sign problem and hence no need for reweighting.
We employ the idea \cite{lom} that in this case
one may fit the non-perturbative data
for an observable by truncated Taylor series in $\mu/T$, thus keeping full control
over the associated systematic error.
In the parameter range where this is possible, and in the absence of non-analyticities,
the series may be analytically continued to real values of $\mu$.
Non-perturbative evidence for the viability of this approach 
has been given for screening masses in the deconfined 
phase \cite{hlp2}, which can be simulated
sucessfully for real and imaginary $\mu$ \cite{hlp1} in
the framework of dimensionally reduced QCD \cite{dr}. 
Here we extend this approach to the critical line $T_c(\mu)$.
This is possible by noting that,
while giving the location of singularities in thermodynamic limit, 
the critical line itself is an analytic function. 
After continuing it from imaginary to real $\mu$ for
various volumes, a finite volume scaling analysis
of its location should also reveal the nature of the
transition. 
Since there is no sign problem, 
there are no limitations on the volumes that can be simulated either.
In particular, it should then be possible to identify the location of 
the critical endpoint of the phase transition which 
is expected on general theoretical grounds \cite{kr},
and for which reweighted numerical results have been presented in \cite{fk2}.

Our method is limited by a maximal imaginary part of the chemical potential, $\mu_I^c=\pi T/3$,
above which the system tunnels into an unphysical $Z(3)$ sector. Along the transition line, 
this corresponds to an accessible range of real $\mu_B\lsim 500$ MeV, which well covers
densities in heavy ion experiments \cite{exp}. 
We present numerical results for the location of the critical line, and
find that the truncation error of the Taylor series is far smaller than
statistical errors in the whole accessible range.
A finite volume scaling analysis is beyond the scope of our current computational
resources and postponed to later work.
Nevertheless, our results indicate that the part of the phase diagram relevant
for heavy ion collisions can be reliably obtained from simulations of
imaginary chemical potentials.

In Sec.~\ref{gen} we recall some general features of QCD at imaginary
chemical potential, which will be useful in the sequel. 
The analyticity of the critical line is derived in Sec.~\ref{ana}. 
Numerical results for two flavor QCD with Wilson gauge action and 
Kogut-Susskind fermions are presented in Sec.~\ref{num}, while Sec.~\ref{comp}
discusses our approach in comparison with previous ones. In Sec.~\ref{con} we
draw our conclusions.

\section{\label{gen} QCD at real and imaginary chemical potential}

While this section contains
no new results, we would like to discuss some qualitative features in detail as they
play an important role for the following.
We denote by $\mu$ the chemical potential
for quark number $Q$, and by  $\mu_B$  the chemical potential
for baryon number $B=Q/3$, i.e.~$\mu = \mu_B/3$. 
The grand canonical partition function is compactly written as
\be
Z(V,\mu,T)=\tr \left(\ex^{-(\hat{H}-\mu \hat{Q})/T}\right).
\ee
An expansion for small chemical potentials
proceeds in terms of the dimensionless parameter $\bar{\mu}=\mu/T$, and hence
our interest is in $Z(V,\bar{\mu},T)$.

The presence of a chemical potential term
in the action breaks the separate invariance under 
Euclidean time reflection and charge conjugation.
In particular, a time reflection can be compensated by $\mu\rightarrow -\mu$ and
vice versa, so that
\be \label{s1}
Z(\bmu)=Z(-\bmu).
\ee
As a consequence, observables symmetric under time reflection are even
functions of $\bmu$.

Let $\mu_R,\mu_I\in \RR$ denote the
real and imaginary parts of $\mu$:
\be
\mu = \mu_R + \ii \mu_I,
\ee
and similarly for $\bmu$.
The qualitative features of SU(N) QCD at imaginary chemical potential, 
$\mu_R=0$, were
studied a long time ago \cite{weiss}.
With dynamical fermions the $Z(N)$ symmetry of the pure gauge theory is 
explicitly broken.
However, in the presence of a complex chemical potential, 
a $Z(N)$ transformation of
the fermion fields is equivalent to a shift in $\mu_I$, thus leading to a
new symmetry:
the partition function is periodic in $\mu_I$ with period $2\pi T/N$ \cite{weiss},
\be \label{s2}
Z(\bmu_R,\bmu_I)=Z(\bmu_R,\bmu_I+2\pi/N).
\ee

\begin{figure}[th]
\vspace*{1cm}
\hspace*{4cm}
\epsfxsize=6cm
\epsffile{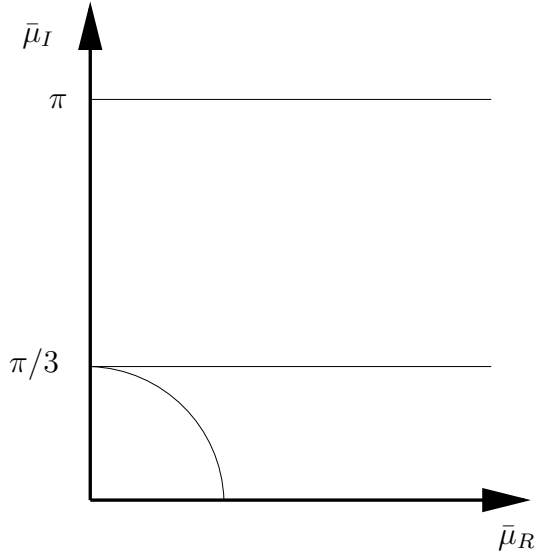}
\put(-15,-15){$\bmu_R$}
\put(-195,175){$\bmu_I$}
\put(-185,150){$\pi$}
\put(-200,50){$\pi/3$}
\caption[]{\label{muri} \em
Location of $Z(3)$ transitions, \eq (\ref{zcrit}).
Analytic continuation is limited to the region within the arc.
}
\end{figure}

The different $Z(N)$ sectors are characterized
by the phase $\varphi$ of the Polyakov loop,
\be
\langle P(x)\rangle = |\langle P(x)\rangle |\ex^{\ii\varphi}.
\ee
At high temperatures one may use the perturbative
effective potential $V(\varphi)$ 
to obtain the qualitative behaviour of the theory.  
In pure gauge theory $\langle P(x)\rangle =0$ for $T<T_c$ and
$\langle P(x)\rangle \neq 0 $ for $T>T_c$. 
In the deconfined phase there are $N$ degenerate minima
at $\varphi=2\pi k/N, k=1,\ldots N$,
separated by potential barriers, and the transition between them
is of first order \cite{zn}.
For $N_f>0, \mu=0$, explicit $Z(N)$-breaking results in an 
expectation value for the Polyakov loop even in the confinement phase. 
The expectation value is real, $\varphi=0$, 
corresponding to the true vacuum of the theory, 
while the other sectors are metastable \cite{meta}. 
This is true both for $T>T_c$ and $T<T_c$.
When a chemical potential is switched on \cite{weiss,sin,laliena}, 
its real part stabilizes this situation
by raising the energy of the other $Z(N)$ vacua, while the imaginary
part has the opposite effect of lowering one of them relative to the $\varphi=0$ vacuum. 
Hence, once $\mu_I$ exceeds some critical value $\mu_I^c$, a phase 
transition to a non-trivial $Z(N)$ sector occurs. 
Non-perturbatively the same statement follows from the aforementioned 
equivalence of certain $\mu_I$-shifts and $Z(N)$ rotations. The combined symmetries
\eqs (\ref{s1}),(\ref{s2}) imply that this transition is periodically repeated
for larger values of $\mu_I$. Moreover, its exact critical values
are dictated to be \cite{weiss}
\be \label{zcrit}
\bmu_I^c=\frac{2\pi}{N} \left(k+\frac{1}{2}\right). 
\ee
On the other hand, 
for purely real $\mu$ this transition
never occurs. This situation is sketched qualitatively for $SU(3)$ in \fig \ref{muri}.
The order of the $Z(3)$ transitions depends on temperature. 
A perturbative calculation
predicts it to be of first order in the deconfined phase, whereas at
low temperature a strong coupling analysis suggests it to be a 
smooth crossover \cite{weiss}. We shall present numerical results in support of
this picture in Sec.~\ref{num}.

Note that the first of these $Z(3)$ transitions at $\bmu_I^c=\pi/3$ limits the
useful information to be obtained from imaginary chemical potentials:
\eqs (\ref{s1}),(\ref{s2}) imply
\be \label{sym}
Z(\bmu_I = \pi/3 + \Delta\bmu_I) = Z(\bmu_I = \pi/3 - \Delta\bmu_I)
\ee
so that all observables are symmetric about the transition at $\bmu_I^c$. 
In other words, all expectation values for
$\bmu_I>\bmu_I^c$ are simply copies of those for $\bmu_I<\bmu_I^c$, determined
by reflection symmetry and periodicity. 
Moreover, here we are interested in
analytically continuing results obtained on the imaginary axis, and this 
is only possible for $|\bmu|\leq \pi/3$ \cite{hlp2}, represented
by the arc of circle in \fig \ref{muri}. 
Furthermore, the
observable to be continued has to be free of additional singularities in this region. 
In the next section we will show that this is the case for the critical
coupling for deconfinement, and hence the associated temperature $T_c(\mu)$. 
Along the deconfinement transition the latter is in the range
$T\sim 160-170$ MeV, which means that our approach is restricted to baryon chemical
potentials $\mu_B\lsim 500$ MeV. 

\section{\label{ana} Analyticity of the (pseudo-) critical line}

On a $L^3\times N_t$ lattice, the rescaled chemical potential takes the form
$\bmu=N_t a\mu$, which up to the constant factor $N_t$ is equivalent to the 
chemical potential in lattice units. On the lattice one then explores the
phase diagram in the $(\beta,\bmu)$-plane, which in the continuum limit
has to be converted to a $(T,\mu)$-diagram.

The location and nature of a phase transition may be determined from
the finite volume scaling of susceptibilities
of dimensionless operators ($V=L^3$),
\be
\chi= VN_t \left\langle(\op - \langle\op\rangle)^2\right\rangle,
\qquad \op=\frac{1}{VN_t}\sum_{\bfx,t}\op(x).
\ee
In pratice we shall use as $\op(x)$ the plaquette, the chiral condensate and the
modulus of the Polyakov loop.
In a transition region, fluctuations are strong and susceptibilities display
a peak, $\chi_{max}=\chi(\bmu_c,\beta_c)$, 
whose location determines the critical parameters.
In a finite volume, the susceptibility is always an analytic function of the parameters
of the theory, even in the presence of a phase transition. The latter 
reveals itself by a divergence of $\chi_{max}$ in the infinite volume limit,
whereas $\chi_{max}$ stays finite in the case of a crossover.
The order of the transition is determined by the rate of divergence:
for a first order transition $\chi_{max}\propto V$, whereas
for a second order phase transition, $\chi_{max}\propto V^{\rho}$
with a critical exponent $\rho = \gamma/d\nu < 1$ \cite{bar}.
Alternatively, one may consider the finite size scaling of the critical coupling
$\beta_c(V)$ itself,  
which attains its infinite volume limit as 
\be
(\beta_c(V)-\beta_c(\infty))\sim V^{-\sigma},
\ee
where $\sigma=1$ for a first order phase transition, $\sigma=1/d\nu < 1$
for a second order phase transition, and $\sigma=0$ for a crossover.
(For notation and numerical application to SU(3) 
pure gauge theory, see e.~g.~\cite{scale} and references
therein).

At zero density, the deconfinement transition is signalled 
by a peak $\chi_{max}=\chi(0,\beta_c)$,
defining the critical coupling through the equations
\be \label{crit}
\left.\frac{\partial\chi}{\partial\beta}\right|_{\beta_c}=0\quad
\left.\frac{\partial^2\chi}{\partial\beta^2}\right|_{\beta_c}<0.
\ee
When the chemical potential is switched on, this peak extends as a
``mountain ridge'' into the $(\beta,\bmu)$-plane. 
Starting at $\beta_c(0)$, the ridge is defined by maximizing $\chi$ while moving 
into the $(\beta,\bmu)$-plane, i.e.~by moving along the direction of the gradient 
$\nabla \chi(\bmu,\beta)$ (or the direction with the largest second derivative
in the case of vanishing gradient). 
It has a vanishing first and negative
second derivative in the direction perpendicular to the gradient. 

\begin{figure}[th]
\vspace*{1cm}
\hspace*{1cm}
\leavevmode
\epsfxsize=6cm
\epsffile{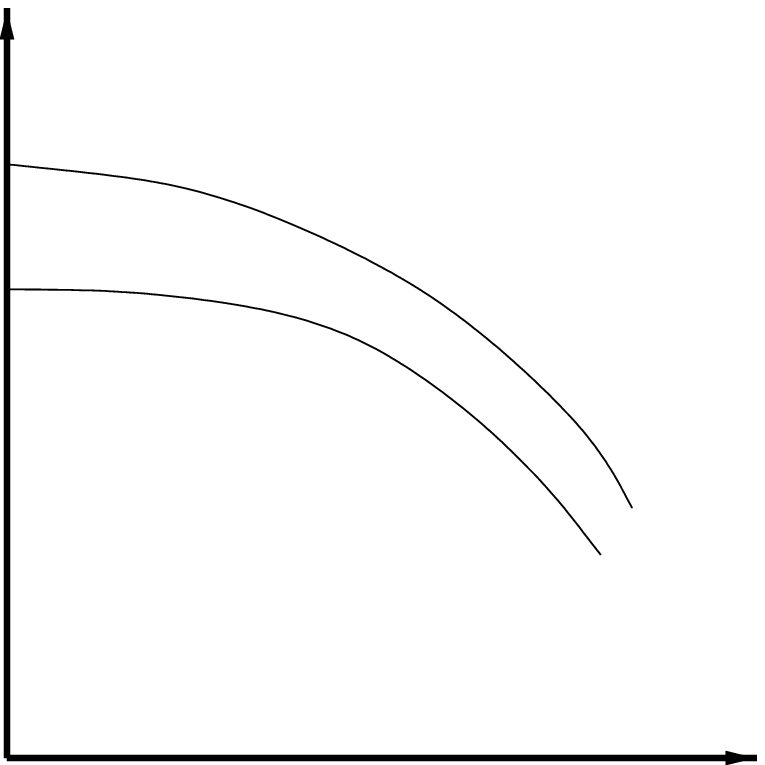}
\put(-15,-15){$\mu$}
\put(-195,160){$T$}
\put(-100,140){\large finite V}
\hspace*{1cm}
\epsfxsize=6cm
\epsffile{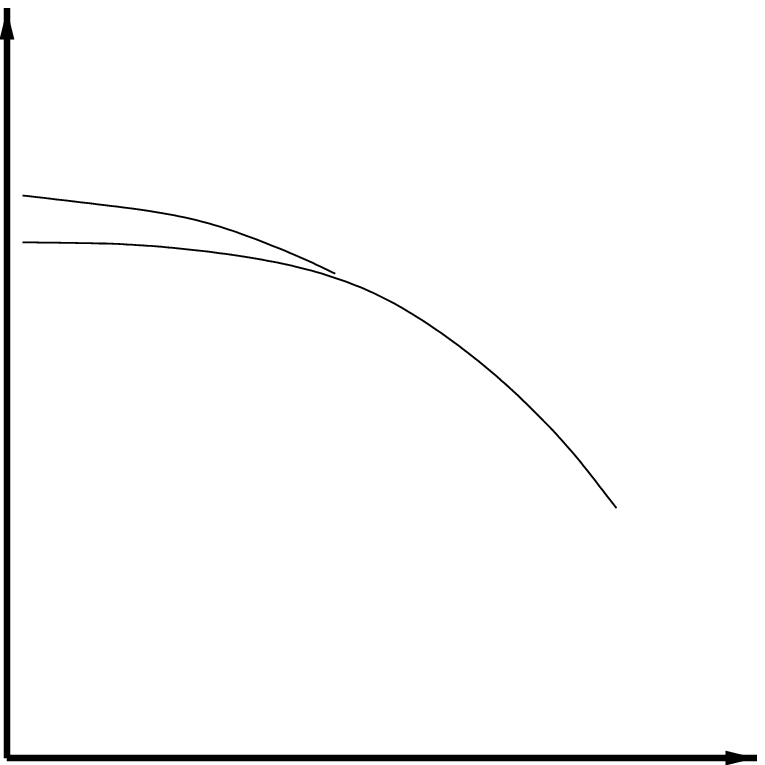}
\put(-15,-15){$\mu$}
\put(-195,165){$T$}
\put(-100,140){\large infinite V}
\caption[]{\label{lines} \em
Schematic location of the lines defined by \eq (\ref{crit1}).
In the infinite volume limit (right), the lines merge for a phase transition, but stay
separate for a crossover, the bifurcation marking the end point of the phase transition.}
\end{figure}

For the numerical analysis, however, 
it is more convenient to consider maxima in the directions of $\beta$ and $\bmu$,
respectively. The two pairs of conditions
\be \label{crit1}
\left\{\partial_\beta\chi =0,\partial^2_\beta\chi<0\right\},\qquad
\left\{\partial_{\bmu}\chi =0,\partial^2_{\bmu}\chi<0\right\}
\ee
then define two separate lines bounding the ridge from below and above, 
as indicated in \fig \ref{lines} (left), 
where the ordering of the lines depends on the sign of the gradient. For large enough 
volumes one expects a positive gradient while increasing $\mu$ towards a critical point, and 
the first condition denotes the lower line. 
When the volume is taken to infinity, the mountain ridge
changes its shape: where there is a phase transition, it becomes singular and
infinitely narrow, i.e.~the two lines merge 
into one. Where there is a crossover,
the ridge saturates at finite height with finite curvature, and the two lines
remain separate (as in \fig \ref{lines} (right) at small $\mu$). In this case there is no 
phase transition and no uniquely defined pseudo-critical line. 
Nevertheless, the physical properties change rapidly in the region
of the line defined by $\partial_\beta\chi=0$, as is well known from the zero density
case. 
Moreover, for small $\bmu$ this line is the one closer to the ridge.
Thus, for every $\bmu$ we may locate the thermal transition by the 
conditions specified in \eq (\ref{crit}), which then represent
our implicit definition of the critical line 
$\beta_c(\bmu)$. 

Let us now consider a finite volume.
Expanding the susceptibility around $\beta_c(0)$ and 
using $\chi(\bmu,\beta)=\chi(-\bmu,\beta)$, it takes the form
\be
\chi(\bmu,\beta)=\sum_{n,m=0}\;c_{nm}\;(\beta-\beta_c(0))^n
\bmu^{2m}.
\label{chi}
\ee
Because $\chi(\bmu,\beta)$ is analytic, its derivatives 
are as well.
In the neighbourhood of the critical line we 
furthermore have $\partial^2_\beta\chi\neq 0$. 
The implicit function theorem\footnote{See e.g. \cite{math}. We thank F.~Wilczek for 
reminding us of this theorem in this context.} then tells us that the implicitly defined
$\beta_c(\bmu)$ is analytic in $\bmu$. 
Furthermore, its first derivative is obtained
from the chain rule for partial differentiation,
\be \label{dbdm}
\frac{\partial\beta_c}{\partial\bmu}=-\frac{\partial^2\chi}{\partial\bmu\partial \beta}
\left(\frac{\partial^2\chi}{\partial\beta^2}\right)^{-1}.
\ee
Since $\chi$ is even in $\bmu$, the same follows from the last equation for $\beta_c$,
which therefore has a Taylor expansion
\be
\beta_c(\bmu)=\sum_{n=0} a_n \bmu^{2n}=\sum_{n=0} c_n (a\mu)^{2n},
\label{beta}
\ee
where we have absorbed the temporal lattice length into the coefficients in the 
second equation, $c_n=a_n N_t^{2n}$.

All of these considerations 
are unchanged if we consider a purely imaginary potential, $\bmu=\ii\bmu_I$.
In this case we may define a corresponding real function
$\tilde{\chi}(\bmu_I,\beta)$ with a Taylor expansion in $\beta,\bmu_I^2$ and
coefficients $\tilde{c}_{nm}$. Clearly, this is just the analytic continuation 
$\tilde{\chi}(\bmu_I,\beta)=\chi(\ii\bmu_I,\beta)$,
with $\tilde{c}_{nm}=c_{nm}(-1)^m$.
Its maximum in the $\beta$-direction is again defining a function
$\beta_c(\bmu_I)$, which by the implicit function theorem is analytic.
The analogue of \eq (\ref{dbdm}) then reads
\be
\frac{\partial\beta_c}{\partial\bmu_I}=-\frac{\partial^2\tilde{\chi}}
{\partial\bmu_I\partial\beta}
\left(\frac{\partial^2\tilde{\chi}}{\partial\beta^2}\right)^{-1}
=-\ii\frac{\partial^2\chi}
{\partial\bmu\partial\beta}
\left(\frac{\partial^2\chi}{\partial\beta^2}\right)^{-1}=
\ii\frac{\partial\beta_c}{\partial\bmu}.
\ee 
Since this equation is true for every $\bmu_I<\bmu_I^c$ we have 
thus established $\beta_c(\bmu_I)=\beta_c(\bmu=\ii\bmu_I)$, i.e.~the critical line
of the deconfinement transition at imaginary chemical potential is simply the 
analytic continuation of the critical line at real chemical potential.

Next, we need to discuss the infinite volume limit.
As remarked above, for every $\bmu$ 
the critical coupling will approach its thermodynamic limit in a fashion characteristic
of the order of the transition. However, $\beta_c(\infty,\bmu)$ is only shifted
from $\beta_c(V,\bmu)$, without developing any singularities in the thermodynamic limit.
Hence, the critical line defined by \eq (\ref{crit})
marks the singularities in the partition function, which it smoothly connects
to a pseudo-critical line
in the crossover region. It thus
remains itself an analytic function of $\bmu$ in the thermodynamic limit for all
$\bmu$.

Our strategy is now to first compute the critical line for imaginary chemical potential
and check the convergence of its Taylor expansion. As we shall see, convergence is surprisingly fast for the whole range of chemical potentials accessible to this method.
Analytic continuation then reduces to simply flipping the sign of the appropriate terms.
Finally, the infinite volume limit has to be taken from the continued results 
at real $\bmu$. The order of the transition can then be determined in a $(V,\bmu)$-range where
the truncation error of the series is smaller than finite size scaling effects.

\section{\label{num} Numerical results for two light flavors}

In order to test the feasibility of our approach, we consider QCD with two flavors
of staggered fermions with bare mass $am=0.025$  and $\mu_R=0$ 
on a $8^3\times 4$ lattice.
We use the R-algorithm~\cite{R-alg} with a step size $\delta\tau = 0.02$,
sufficiently small for the systematic errors ${\cal O}(\delta\tau^2)$ to be
negligible compared to our statistical errors. For 8 values of $\beta$ 
spanning the relevant temperature regime and 6 values of $a\mu_I\in \{0,\pi/12\}$, 
we have accumulated $2000-6000$ unit-length trajectories each, 
measuring
the gauge action, the Polyakov loop and the chiral condensate after each
trajectory. 
For one parameter set we have performed additional simulations on $6^3\times4$.
To determine the pseudo-critical value $\beta_c(a\mu_I)$, we use the
Ferrenberg-Swendsen reweighting method~\cite{FS}.

\subsection{\label{zt} The Z(3) transition for imaginary $\mu$}

\begin{figure}[th]
\vspace*{1cm}
\centerline{\epsfxsize=8cm\hspace*{0cm}\epsfbox{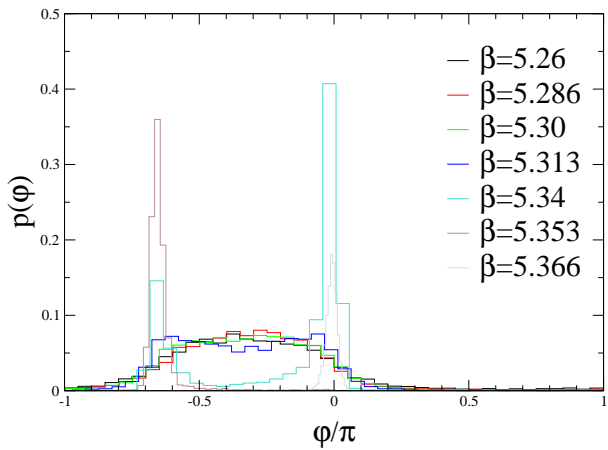}}

\caption[a]{\em 
Probability distribution of the phase of the Polyakov loop for the critical
value $a\mu_I^c=\pi/12$.
}
\label{phaseP_hist}
\end{figure}

First, it is instructive to investigate the 
Z(3) transition discussed in Sec.~\ref{gen}. 
The predicted critical value for the chemical potential is at
$\bmu_I^c=\pi /3$, which on our $N_t=4$ lattice corresponds to 
$a\mu_I^c = \pi/12=0.262$. 
For this value 
\fig \ref{phaseP_hist} shows histograms
of the phase $\varphi$ of the Wilson loop, obtained for various lattice couplings
(temperatures). We find the system tunneling between the sectors with $\varphi=0$
and $\varphi=-2\pi/3$.
\begin{figure}[ht]
\vspace*{1cm}
\leavevmode
\epsfxsize=7.5cm\hspace*{0cm}\epsfbox{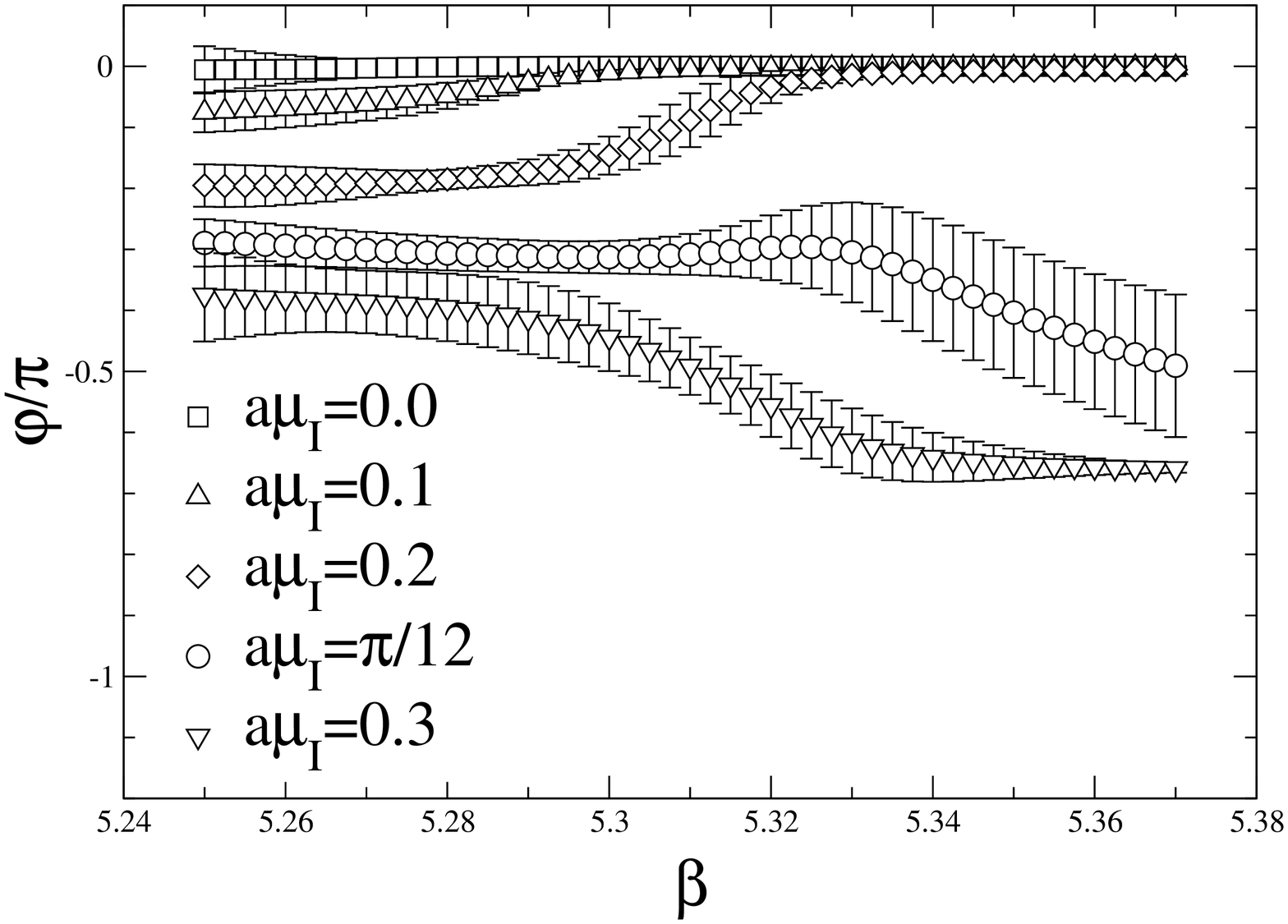}
\epsfxsize=7.5cm\hspace*{0.5cm}\epsfbox{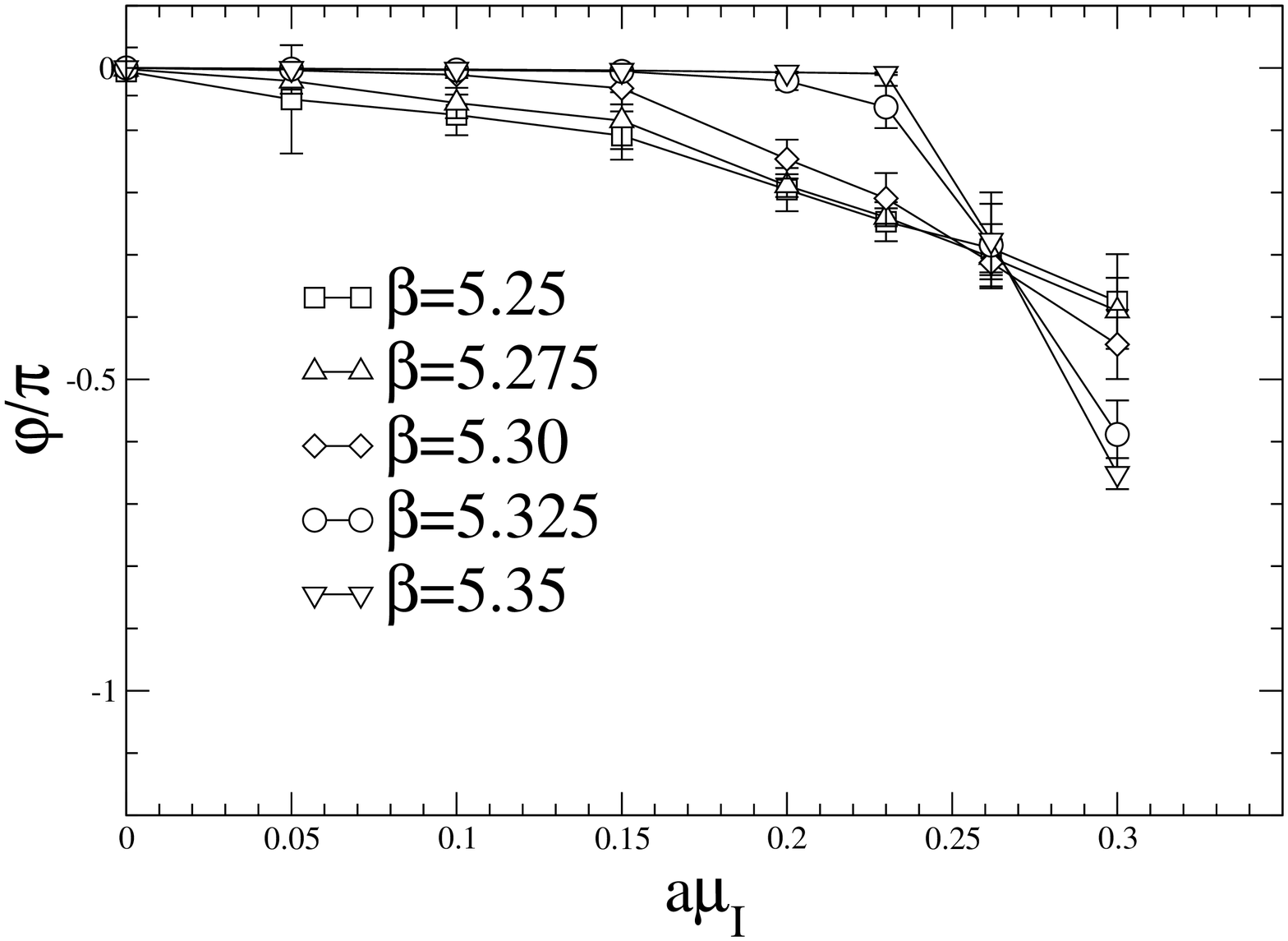}

\caption[a]{\em
The $\beta$- and $\mu_I$-dependence of the average phase of the Polyakov loop.
}
\label{phaseP}
\end{figure}
\begin{figure}[th]
\vspace*{1cm}
\centerline{\epsfxsize=8cm\hspace*{0cm}\epsfbox{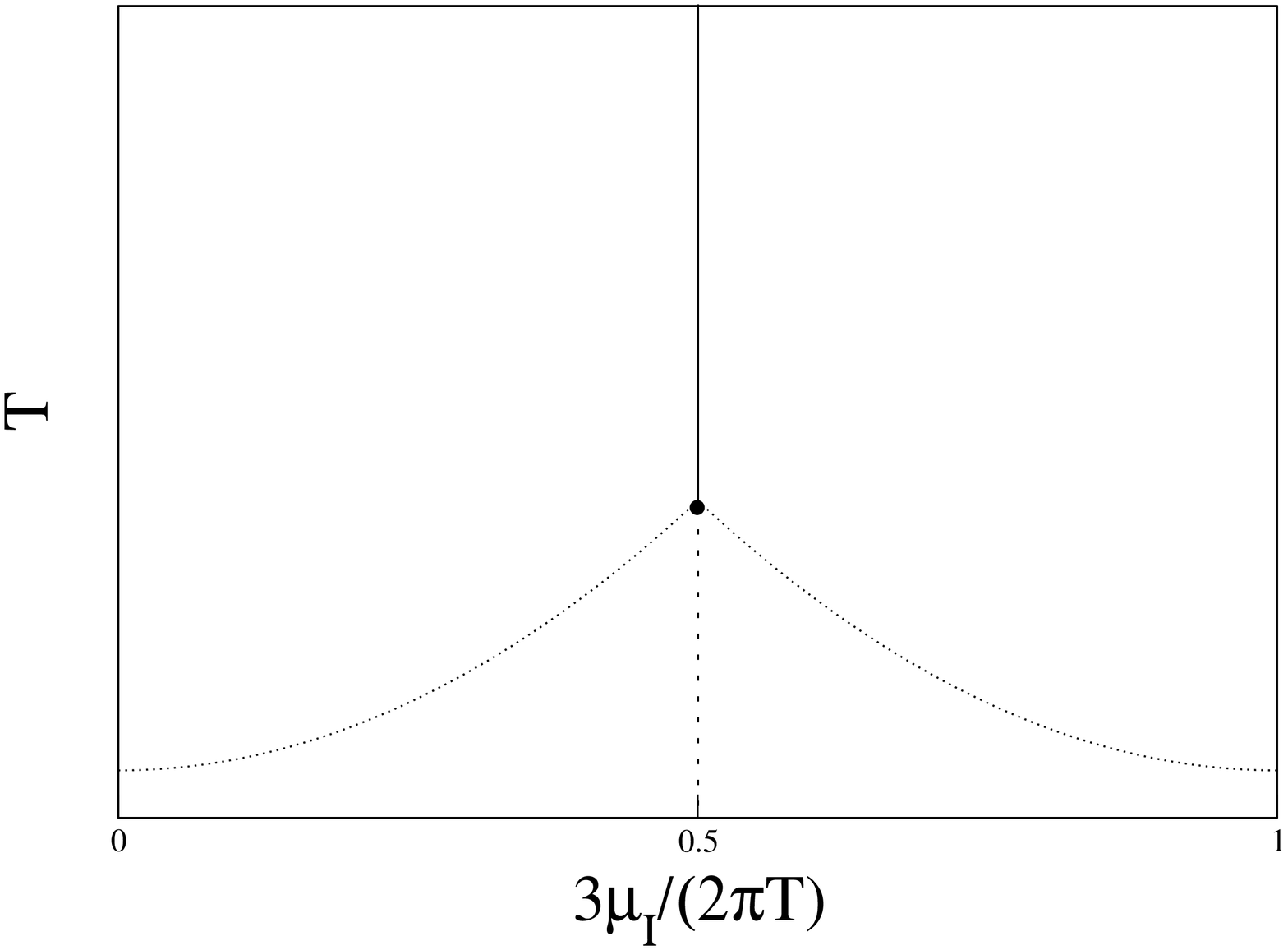}}

\caption[a]{\em
Schematic phase diagram for imaginary chemical potential:
The vertical line marks the $Z(3)$ transition (this section), the curved lines
the deconfinement transition (next section). 
The solid line indicates a first order transition, while the 
nature of the dotted lines is not yet determined. The diagram is periodically repeated 
for larger values of $\mu_I$.
}
\label{schem}
\end{figure}

In accord with the predictions made in \cite{weiss}, 
we observe a smooth distribution consistent with a crossover at low temperatures,
and a pronounced two-state signal indicating a first order phase transition
at high temperatures. The critical coupling for the onset of
the $Z(3)$ transition is visibly between $5.313<\beta_c<5.34$.

For low temperatures, where the $Z(3)$ transition is continuous,
there is frequent tunneling between the $Z(3)$ sectors.
It then follows that also for $a\mu_I<(a\mu_I)^c$ 
the ensemble average consists of a mixture of the two sectors, where the 
$\varphi=-2\pi/3$ admixture grows with $a\mu_I$. This is clearly visible in 
\fig \ref{phaseP}, where the average phase of the Polyakov loop gradually moves from
$\varphi(a\mu_I=0)=0$ to $\varphi((a\mu_I)^c)=-\pi/3$, the latter value corresponding
to the average between the two phases $\varphi=0,-2\pi/3$. For $a\mu_I>(a\mu_I)^c$,
the phase gets dominated by the vacuum with $\varphi=-2\pi/3$.

At high temperatures, on the other hand, the $Z(3)$ sectors are separated by a 
diverging potential barrier, and tunneling 
is more suppressed the higher the temperature. 
As a result the Monte Carlo ensemble stays longer in
one sector, and its equilibrium is dominated by the lowest vacuum. Thus, 
as shown in \fig \ref{phaseP}, for $a\mu_I<(a\mu_I)^c$ the system always settles in the
trivial vacuum with zero phase, whereas for $a\mu>(a\mu_I)^c$ it settles in the vacuum 
sector with $\varphi=-2\pi/3$. The dividing line between those situations 
is $\varphi((a\mu_I)^c)=-\pi/3$, which is independent of temperature by symmetry.
In \fig \ref{phaseP}, one observes that indeed $\varphi((a\mu_I)^c)=-\pi/3$ for
all $\beta$'s, although errors become large at large $\beta$. This reflects
the difficulty of tunneling at high temperatures, and implies longer and longer
Monte Carlo time for the system to equilibrate. 
Hence, at high temperatures great attention must be paid to $Z(3)$ ergodicity
and its possible violation.

\begin{figure}[pth]
\vspace*{1.0cm}
\centerline{\epsfxsize=7cm\hspace*{0cm}\epsfbox{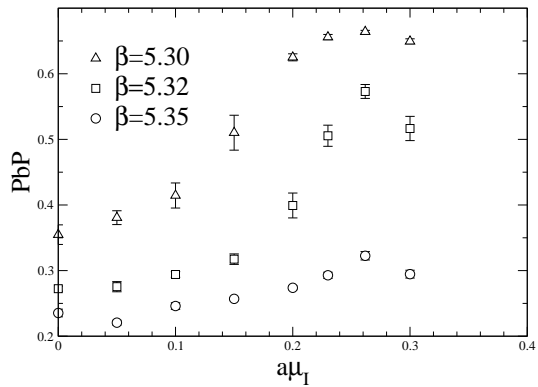}}

\caption[a]{\em
The chiral condensate $\bar{\psi}\psi$ as a function of $a\mu_I$.
}
\label{PbPmu}
\end{figure}

From the susceptibilities $\chi(\beta)$ at $(a\mu_I)^c = \pi/12$, only one
peak is apparent. This means that,
within our accuracy,
at $(a\mu_I)^c$ the critical temperature
at which the $Z(3)$ first-order transition line starts
is the same as the critical temperature for the deconfinement transition
determined below.
Schematically the $(T,\mu_I)$ phase diagram then looks as in \fig
\ref{schem}, with a special point where three critical lines meet.
This point might be tricritical if deconfinement corresponds to a genuine
phase transition, or critical otherwise. We take a first look at this issue below.

Once ergodicity is ensured, the effects of the Z(3) transition are also
visible in other observables like, e.g., the chiral condensate, as shown in
\fig \ref{PbPmu}. Note the symmetry of the observable around $a\mu_I^c$ according to
\eq (\ref{sym}). For low $\beta$ (viz.~temperatures) the observable 
is continuous at the critical chemical potential, while a cusp indicating 
a discontinuous transition develops for large $\beta$.
\begin{figure}[b]
\vspace*{1cm}
\leavevmode
\epsfxsize=7.5cm\hspace*{0cm}\centerline{\epsfbox{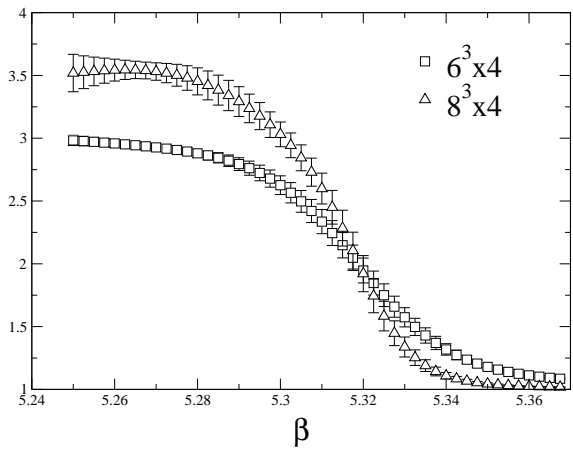}}
\caption[a]{\em
Ratio of of cumulants $\langle \hat\varphi^4 \rangle/\langle \hat\varphi^2 \rangle^2$,
for $a\mu_I=\pi/12$.
}
\label{cumulant}
\end{figure}

Let us now try to 
determine the endpoint $(\beta_c,a\mu_I = \pi/12)$ of the 
first-order transition line with better accuracy. 
We can do so by monitoring the plaquette susceptibility
as a function of $\beta$. This procedure gives us $\beta_c = 5.325(5)$.
For improved accuracy, and to check our control of the delicate ergodicity
problems mentioned above, we performed additional simulations on a $6^3\times 4$
lattice at $a \mu_I = \pi/12$. Since $\varphi = -\pi/3$, one can form a ratio 
of cumulants 
$\langle \hat\varphi^4 \rangle/\langle \hat\varphi^2 \rangle^2$,
where $\hat\varphi = \varphi + \frac{\pi}{3}$, and estimate the critical
point $\beta_c$ from the crossing of the cumulant ratios for the two 
available volumes. 
This procedure, illustrated in \fig\ref{cumulant},
yields a consistent value $\beta_c = 5.321(7)$.
Note also that the crossing occurs at a value of the cumulant ratio
$1.8(2)$ consistent with that of the $3d$ Ising universality class
$1.604(1)$~\cite{ising}. 
This indicates as the simplest interpretation 
of the phase diagram: the first-order $Z(3)$ transition line ends with
a second-order critical point in the Ising class. This picture will likely
depend on the number of quark flavors and on their masses.
\begin{figure}[hb]
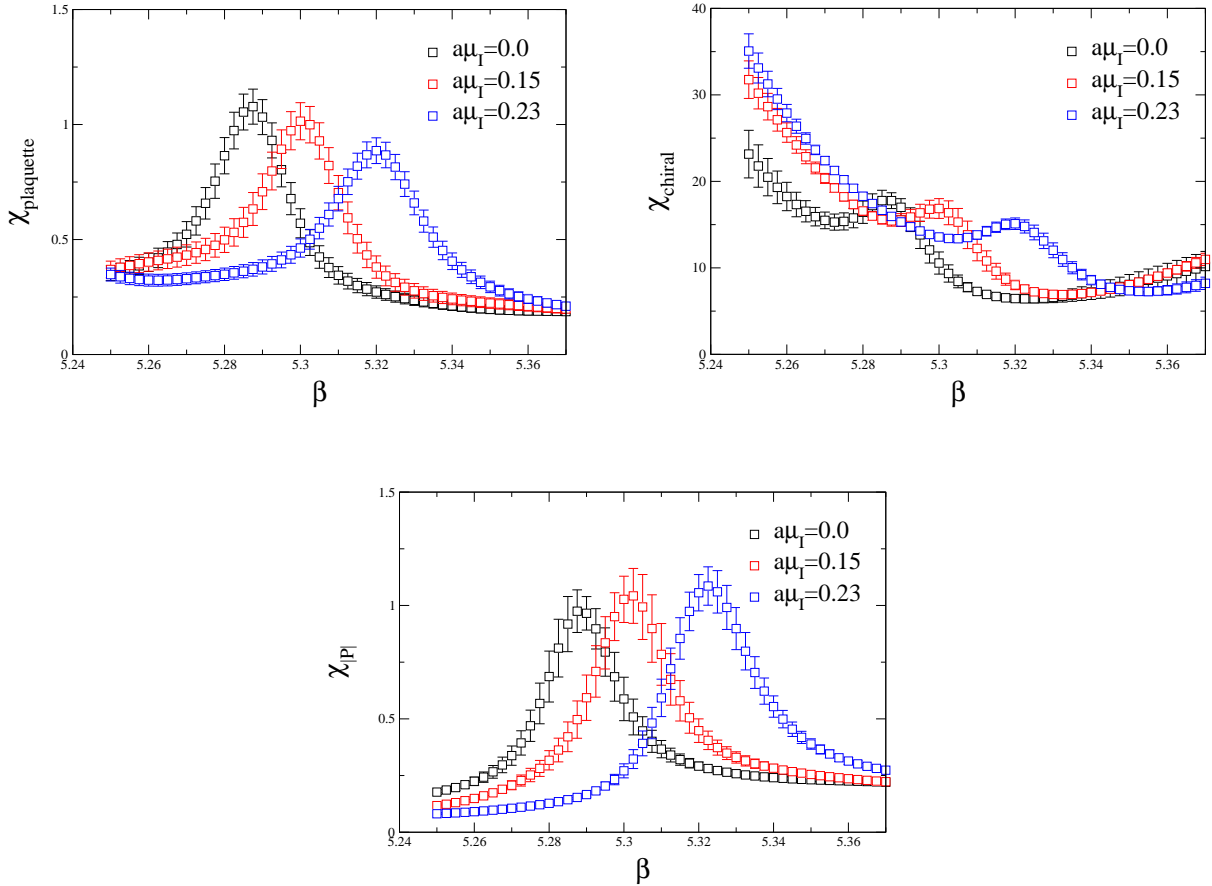

\vspace*{1cm}
\centerline{\epsfxsize=7.5cm\hspace*{0cm}\epsfbox{plaqsus_all.eps}%
            \epsfxsize=7.5cm\hspace*{1cm}\epsfbox{PbPsus.eps}}
\vspace*{1.0cm}
\centerline{\epsfxsize=7.5cm\hspace*{0cm}\epsfbox{absPsus.eps}}

\caption[a]{\em
Susceptibilities for the plaquette, the chiral condensate and the absolute value
of the Polyakov loop.
}
\label{susc}
\end{figure}
\begin{figure}[t]
\vspace*{1.0cm}
\centerline{\epsfxsize=8cm\hspace*{0cm}\epsfbox{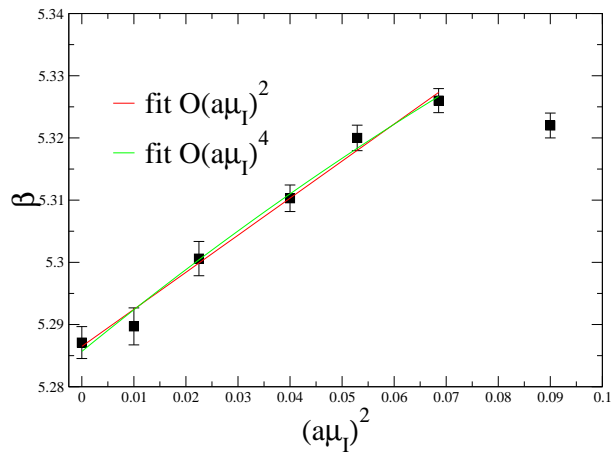}}

\caption[a]{\label{pdiag}\em
Location of the pseudo-critical deconfinement line as determined from plaquette
susceptibilities. The lines show the fits given in Table \ref{tab}.
}
\end{figure}
\newpage
\subsection{The deconfinement transition for imaginary $\mu$}

In order to map out the deconfinement 
transition line in the $(\beta,\bmu_I)$-plane, we have measured
the susceptibility of the plaquette, of the modulus of the Polyakov loop, and
the disconnected part of a stochastic estimator for the chiral condensate susceptibility. 
All three observables yield consistent 
results for $\beta_c(a\mu_I)$, as shown
in \fig \ref{susc}. We have chosen to display only three $a\mu_I$-values for clarity
of the plots. 
Note that the critical lattice coupling is growing 
with increasing $a\mu_I$, and thus the critical temperature is an increasing function
of $\mu_I$ as well. This is in contrast to the situation at real $\mu$, where
$T_c(\mu)$ is a decreasing function. This qualitative behaviour is in accord
with the one found for screening lengths in units of inverse temperature, 
which are increasing (decreasing) functions
of $\mu_I$ ($\mu_R$) \cite{hlp2}. (An increasing screening length means more energy
is required to separate charges before they are screened, and hence an increasing 
deconfinement temperature).

\begin{table}[tb]
\begin{center}
\begin{tabular}{|c|*{4}{r@{.}l|}l|}
\hline
\hline
$\beta_c$ from $\chi_{plaq}$ &
\multicolumn{2}{c|}{$c_0$} &
\multicolumn{2}{c|}{$c_1$} &
\multicolumn{2}{c|}{$c_2$} &
\multicolumn{2}{c|}{$\chi^2/{\rm dof}$} &
 $Q$ \\
\hline
 $\op\left((a\mu_I^2)\right)$ & 5&2865(18) & 0&596(40) & \none & 0&60 & 0.66 \\
 $\op\left((a\mu_I)^4\right)$ & 5&2857(23) & 0&68(15) & $-1$&2 (2.0) & 0&67 & 0.57 \\
\hline
$\beta_c$ from $\chi_{|\langle P\rangle |}$ &
\multicolumn{2}{c|}{$c_0$} &
\multicolumn{2}{c|}{$c_1$} &
\multicolumn{2}{c|}{$c_2$} &
\multicolumn{2}{c|}{$\chi^2/{\rm dof}$} &
 $Q$ \\
\hline
 $\op\left((a\mu_I)^2\right)$ & 5&2874(17) & 0&640(38) & \none & 0&29 & 0.89 \\
 $\op\left((a\mu_I)^4\right)$ & 5&2873(21) & 0&65(14) & $-0$&2 (2.0) & 0&38 & 0.77 \\
\hline
\hline
\end{tabular}
\caption{ \label{tab}
  {\em
The first coefficients of the Taylor expansion of the critical coupling,
\eq (\ref{beta}). Data and fits for $\beta_c$ from $\chi_{plaq}$
are shown in \fig \ref{pdiag}.
}}
\end{center}
\end{table}

Since the plaquette susceptibility is the one that is most accurately determined,
it is our choice for determining the critical line.
The values of $\beta_c$ corresponding to six values of
$a\mu_I$ are shown in \fig \ref{pdiag}.
Our next task is to determine the reliability of the Taylor expansion for the
critical line, \eq (\ref{beta}). Results of fitting the data to a polynomial
of degree one and two in $(a\mu_I)^2$, respectively, are shown in Table \ref{tab}
and \fig \ref{pdiag}.
Excellent fits are obtained in both cases with  compatible coefficients
$c_{0,1}$.
The quartic correction appears negligible even near $a\mu_I=(a\mu_I)^c$, leaving
$c_2$ to be consistent with zero within this range.
This is again in accord with the analogous finding for screening masses, which are
equally well described by the leading $(a\mu_I)^2$ term in that range \cite{hlp2}.
Statistically consistent results are obtained when $\beta_c$ is extracted from
Polyakov loop instead of plaquette susceptibilites, as also shown in Table \ref{tab}.
We then conclude that
the (pseudo-) critical line
for imaginary chemical potential $\bmu=\ii\bmu_I$, $\bmu_I\leq\pi/3$,
is well described by
\be
\beta_c(a\mu_I)=c_0+c_1(a\mu_I)^2,
\ee
with $c_0, c_1$ as per Table \ref{tab} (top line).

\subsection{The deconfinement transition in physical units for real $\mu$}

With the results of the previous section, it is now trivial to obtain
$\beta_c(\bmu)$ for real $\bmu$, by continuing $\bmu_I\rightarrow -\ii\bmu_I$. 
As a comment on systematics, let us add that the difference between the linear and
quadratic fits in $(a\mu_I)^2$ from Table \ref{tab} gets slightly amplified
by analytic continuation (only the sign of one term is flipped). This reflects the fact
that convergence properties in the real and imaginary directions may 
be different in general. Such an effect will be large for an observable with pronounced
structures, and small on a smooth observable like ours. In any case, the effect can
be monitored, and the error band of the first fit is even after 
continuation completely contained
within the error band of the second fit, which can hence be dropped as statistically 
insignificant.

It is instructive to translate our result into physical units, in order to illustrate
the phenomenological relevance of our approach. 
However, we caution that at this stage such 
a conversion is merely illustrative, with quantitative numbers still afflicted by
various systematic errors: while we have monitored that the truncation error 
of the Taylor series is negligible,
there might still be sizeable corrections from 
finite volume and lattice spacing effects, and we have not investigated the
quark mass dependence of our results. 
\begin{figure}[ht]
\vspace*{1.0cm}
\centerline{\epsfxsize=8cm\hspace*{0cm}\epsfbox{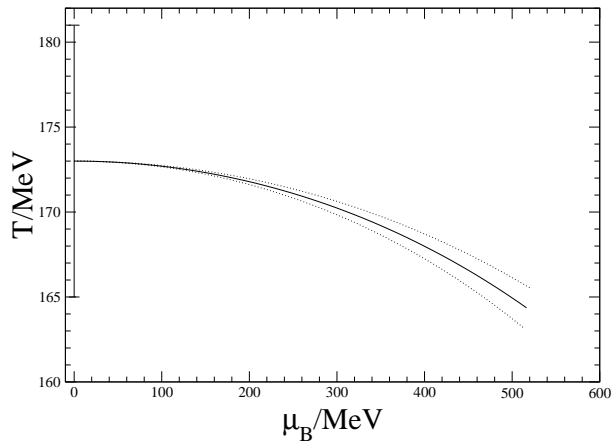}}

\caption[a]{\em
Location of the deconfinement transition corresponding to the
first fit in Table \ref{tab}. The error bar gives the uncertainty
in $T_c(0)$ used to set the scale, the dotted
lines reflect the error on $c_1$ from Table \ref{tab}.
}
\label{rpdiag}
\end{figure}
In view of this we content ourselves
with the perturbative two-loop expression for the lattice QCD scale 
$\Lambda_L$ in the presence of two massless flavors,
\ba
a(\beta)\Lambda_L&=&\left(\frac{6b_0}{\beta}\right)^{-b_1/2b_0^2}\exp(-\beta/12b_0),\nn\\
b_0&=&\frac{1}{16\pi^2}\left(11-\frac{2}{3}N_f\right),\quad
b_1=\left(\frac{1}{16\pi^2}\right)^2\left(102-\frac{38}{3}N_f\right),\nn\\
\frac{T_c(\mu)}{T_c(0)}&=&\frac{a(\beta_c(\mu))\Lambda_L}{a(\beta_c(0)\Lambda_L}
\ea
while the scale is set by the critical temperature $T_c(\mu=0)=173(8)$MeV 
in the chiral limit for staggered fermions \cite{kar} 
(which is expected to still
contain finite lattice spacing errors). 
This leads to the result for the (pseudo-) critical line as shown in
\fig \ref{rpdiag}. 
As remarked before, a detailed finite volume study is beyond the
scope of the present work but would reveal the order of the transition along the line.

\section{\label{comp} Comparison with other methods}

Before concluding, we would like to make some comments comparing our method
with previous studies of the finite $\mu$ deconfinement 
transition \cite{fk2,hk}. The main disadvantage of our approach
is its limitation to the range $|\mu|/T<\pi/3$, as discussed
in Sec.~\ref{gen}. On the other hand, within this range we have full control
over the necessary approximation, i.e.~the truncation of the Taylor series,
which we find to pose no limitation at all within the accessible range. 

We would like to illustrate the importance of this aspect by a comparison with the 
reweighting method \cite{fk1,fk2}. In Ref.~\cite{fk1}, the overlap
of the reweighted with the full ensemble is tested for imaginary $\mu$.
The configurations used for reweighting from the $\mu=0$ ensemble 
are all in the real $Z(3)$ sector, so that reweighting is by construction 
insensitive to the $Z(3)$ transition. This is clearly seen in 
\fig 1 of Ref.~\cite{fk1}, where the reweighted chiral condensate is found
to be a monotonically increasing function of $(a\mu_I)$ in the range $0\leq a\mu_I\leq 0.32$.
Such a result is inconsistent with the symmetries of the imaginary-$\mu$ theory.
Ref.~\cite{fk1} works at the same
lattice spacing as we do, and therefore the critical chemical potential for
$Z(3)$ tunneling is at $(a\mu_I)^c=\pi/12$.
According to \eq (\ref{sym}),
the chiral condensate has to be symmetric about this point and decrease for
larger $a\mu_I$, as it does in our data, \fig \ref{PbPmu}.
Not surprisingly, reweighting completely misses this qualitative behavior\footnote{\fig 1 
of Ref.~\cite{fk1} also shows the chiral condensate rising monotonically
in the {\em full} ensemble. A possible reason for this strange behavior might be
insufficient thermalization, cf.~the discussion in Sec.~\ref{zt}.
Indeed, at yet larger $\mu_I$, the authors of Ref.~\cite{fk1} find, as expected,
clear disagreement between reweighted and full ensemble results \cite{ZF_private}.}.
In the case of real $\mu$, there is no $Z(3)$ transition.
However, the relevance of complex $Z(3)$ sectors was studied in \cite{laliena}.
It was argued there that constraining the Polyakov loop to be real is a 
reasonable approximation below and above the deconfinement transition, 
but a poor one near it.
In any case, the discussion illustrates a potential danger of any 
reweighting method: it may
completely miss a qualitative change in the physics without announcing its failure,
which only shows up by comparison with results obtained without reweighting.

Ref.~\cite{fk2} studied $N_f=2+1$ flavors, with light quark mass values
and lattice spacing as ours, whereas the bare strange quark mass was chosen
eight times heavier than the light ones. Keeping this difference in mind,
we find the location of our continuum transition line consistent with theirs over
the whole range we have studied.

As mentioned in the introduction, another recent work \cite{hk} employs a 
Taylor expansion of the reweighted path integral. 
However, in this case only
susceptibilities are measured, so that one has no control over the higher order
terms. 
Let us recall that the Taylor expansion is in $\mu/T$, which
becomes larger than one already at $\mu\sim 170$ MeV. A priori there is no way
of knowing how large the convergence radius of such an expansion is.
Simulating at imaginary $\mu$ provides a framework for obtaining quantitative
information about the convergence. As it turns out, our results non-perturbatively 
endorse the use of the leading Taylor coefficient in $T_c(\mu)$ 
up to $\mu\sim 170$MeV. 
Ref.~\cite{hk} works at quark masses four to ten times larger than ours, but
finds only very weak quark mass dependence.
Keeping this difference in mind,
we find our result for the slope $c_1$ of $\beta_c((a\mu)^2)$ to be consistent
with theirs.

\section{\label{con} Conclusions}

The past year has seen progress in the numerical study of finite 
density QCD. However, this progress is essentially based on adopting a
more pragmatic viewpoint. The net quark density in heavy-ion collisions
is small. Therefore, instead of solving the fundamental difficulty posed
by the sign problem, it is very useful already to explore the regime of
small chemical potential, where the sign problem is moderate and various
approaches appear possible.

We have studied QCD with two light flavors of staggered fermions
in the presence of an imaginary chemical potential $\mu=\ii\mu_I$.
In this case the partition function is periodic in $\mu_I$ 
and, in addition to the deconfinement transition,
$Z(3)$ transitions occur at $(\mu_I/T)_c=2\pi(n+1/2)/3$.
We mapped out the phase diagram in the $(T,\mu_I)$-plane
and found that the $Z(3)$
transitions are of first order in the deconfined phase and continuous
in the confined phase.
For $(\mu_I/T)<\pi/3$, the critical temperature for deconfinement is an
analytic function, irrespective of the order of the transition.
Within this range its Taylor expansion is found to rapidly converge and
to be well described by the leading term.

Other recent work uses reweighting of simulations performed at $\mu=0$,
and at one or more temperatures. In contrast, we vary two parameters in our
simulations: the temperature as usual, and also the imaginary chemical
potential $\mu_I$. Having at our disposal two-dimensional information
allows us a reasonable control over systematic errors.

We have used this property to determine the transition line $T_c(\mu)$
for real $\mu$ through analytic continuation.
While we cannot, on a single volume, find out whether the transition is
a crossover or a genuine phase transition at each value of $\mu$, 
a finite-size scaling study should elucidate this issue, provided that finite size
effects are larger than the truncation error. Our central
result is shown in Fig.~\ref{rpdiag}. 
It is consistent with Refs.~\cite{fk2,hk}. The transition line is well
represented by the equation
\be
\frac{T_c(\mu_B)}{T_c(\mu_B=0)}= 1 - 0.00563(38) \left(\frac{\mu_B}{T}\right)^2,
\ee
while the next-order term $\op((\mu_B/T)^4)$ is statistically insignificant
up to $\mu_B \sim 500$ MeV.
\\

\noindent
{\bf \large Acknowledgements:} We thank V.~Azcoiti, O.~B\"ar, Z.~Fodor,
M.~Garc\'{\i}a~P\'erez, M.~Laine, M.-P.~Lombardo, K.~Rajagopal and F.~Wilczek 
for discussions and comments.

\end{document}